\documentclass[amsmath,amssymb,aps,prb,superscriptaddress,twocolumn,showkeys]{revtex4-1}

\usepackage{latexsym,amsbsy,amsfonts,graphicx}
\usepackage{hyperref}
\usepackage{epstopdf}
\usepackage[utf8]{inputenc}
\usepackage{color}
\usepackage{verbatim}
\usepackage[english]{babel}

\usepackage{color}
\usepackage{bm}

\newcommand{\Tr}{\mathop{\rm Tr}\nolimits}

\newcommand{\e}{\mathop{\rm e}\nolimits}
\newcommand{\vecr}{\vec{r}}

\begin{document}

\title{Graphene under the influence of Aharonov--Bohm flux and constant magnetic field
}
\author{E. A. Stepanov}
\affiliation{Radboud University, Institute for Molecules and Materials, 6525AJ Nijmegen, The Netherlands
}

\author{V. Ch. Zhukovsky}
\affiliation{Faculty of Physics, Moscow State University, 119991,
  Moscow, Russia
}

\date{\today}

\begin{abstract}
Investigation of real two--dimensional systems with Dirac--like
electronic behavior under the influence of magnetic field is challenging and leads to many interesting physical results. In this paper we study 2D graphene model with a particular form of magnetic field as a superposition of a homogeneous field and an Aharonov--Bohm vortex. For this configuration, electronic wave functions and energy spectrum were obtained and it was shown that the magnetic Aharonov--Bohm vortex plays the role of a charge impurity. As a demonstration of vacuum properties of the system, vacuum current, as well as an electric current, is calculated and their representation for particular limiting cases of magnetic field is obtained.     
\end{abstract}



\maketitle

\section{Introduction}
Recent investigations in condensed matter physics revealed a number of materials, that can be described by the effective two dimensional Dirac equation. The well known example of such systems is graphene, the planar monoatomic layer of carbon \cite{Novos, Geim10, Kats, Ge, Geim0}, where the gapless nature is protected by sublattice symmetry. Band structure of the $p_{z}$ orbitals in graphene is formed by the two bands with the linear dispersion at the Fermi level at two inequivalent points of the Brillouin zone. The nearest--neighbour hoppings in graphene are much more stronger, than next to nearest--neighbor ones, thus electrons can be effectively modelled by using a continuum version of the tight-binding model, leading to the Dirac equation for massless fermions \cite{wallace, Semenoff, Gus, Neto2}. 

Further researches of electronic behavior in 2D models by the Dirac equation with account for topological properties were made in \cite{Hou, Chamon, Obispo, Obispo1}. In particular, the specified study in the application to graphene with different types of defects \cite{Lahiri, Huang, Ott} gave important results concerning nontrivial properties of transmission of propagating particles. Note also the recent theoretical studies of electronic transport through line defects in graphene \cite{China2, line, Gun1, Gun2}, which can be used to control the electronic transport in graphene. Furthermore, the electrons in the surface states of the 3D topological insulators \cite{Ti2, Ti3, Ti1}, where the crossing point is also protected by topology, and in 3D Weyl semimetals \cite{Wsm1, Wsm3, Wsm2} behave as two--dimensional Dirac particles as well. Therefore, investigations of the 2D materials with nontrivial topology with the use of the Dirac equation are of great importance nowadays.  

On the other hand, applying of the electric field to Dirac systems is also challenging and leads to many important features. For example, recent investigations in graphene and graphene like models under the influence of AC \cite{Ac1, Ac2, Ac3} and DC \cite{Novos, Dc2} electric fields demonstrate a possibility to change the band structure of the system and the dynamics of quasiparticles. Magnetic fields of complicated configuration might also be the cause of various non-trivial effects. Exact solutions of relativistic wave equations for the 3D and 2D Dirac systems with the superpositions of Aharonov--Bohm, magnetic, and electric fields  were obtained in \cite{BGT,BGT1}. Further studies of the effect of the Aharonov--Bohm magnetic field, produced by a thin solenoid in plane structures, were made in \cite{JaMi, PhysRevB.83.075420, Slob1, Slob2} and resulted in calculations of the density of states, induced charge density and induced current. It is worth mentioning, that similar effect has been investigated in \cite{PhysRevD.60.125017, SITENKO2000167} in the context of quantum field theory. 

In this paper we develop  a continuum description of the Dirac model of graphene with the superposition of a constant and homogeneous magnetic field and an Aharonov--Bohm vortex of finite radius. We obtain the wave functions and energy spectrum of electrons and show that the Aharonow--Bohm vortex in graphene plays the role of a charged defect, which certainly affects the energy spectrum. We also derive, using the effective potential method, the vacuum current induced in the model, and consider the electric current in the strong magnetic field limit and show, that it is directed along the Aharonov--Bohm vortex.   

The paper is organized as follows. In Section II we introduce the Dirac equation for the system with the homogeneous magnetic field and the Aharonov--Bohm vector potential. Upon solving the equation of motion it was possible to obtain the eigenvalues and eigenstates of the problem and show that the vortex field affects the spectrum as the charged defect only in the case, when it is captured by the electronic trajectory. In Section III we introduce the effective potential and study the weak and strong field limit. Vacuum current is also obtained in this section for the both limits of magnetic field strength. Section IV is dedicated to calculation of the electric current in the strong magnetic field limit and finally, Section V contains a summary and conclusions.  

\section{Dirac equation}
We consider the magnetic field configuration to consist of two separate components: a homogeneous magnetic field and an Aharonov--Bohm field. In cylindrical coordinates $(r,\,\varphi,\,\,z)$ a homogeneous magnetic field is directed along the $z$ axis ($\vec B || Oz$), which is perpendicular to the graphene surface. The vector potential of the Aharonov--Bohm field has only angular component $\vec{A}^{AB}=(0,\,\,A_{\varphi}^{AB},\,\,0)$ and is regularized by the finite radius of the circle $R$. Then, the vector potential of the total magnetic field configuration can be written as
\begin{equation}\vec{A}(r)=A_{\varphi}\vec{e}_{\varphi}=
\left(A_{\varphi}^{AB}\theta(r-R)+
\frac12 Br\right)\vec{e}_{\varphi},
\end{equation}
where $A_{\varphi}^{AB}$ is the Aharonov--Bohm potential 
\begin{equation}\label{AB}
A_{\varphi}^{AB}=\frac{\Phi}{2\pi{}r}
\end{equation}
with $\Phi=\mu\Phi_{0}$ ($\mu =~ {\rm const}$) as the total magnetic flux of the vortex, and 
\begin{align}
\Phi_{0}=2\pi\hbar{}c/|e|
\end{align}
is the magnetic flux quantum of the electron (it is assumed that $\hbar=c=1$ in what follows due to the appropriate choice of system of units). This vector potential produces the total magnetic field of the form 
\begin{equation}
\vec{B}(r)=\vec\nabla\times{}\vec{A}=\left(B+\frac{\Phi}{2\pi{}R}\delta(r-R)\right)\vec{e}_{z}.
\end{equation}

It should be mentioned here, that in our problem we consider the solution in the region $r>R$ only, due to the fact, that the radius of the Aharonov--Bohm vortex $R$ is usually small enough and comparable to the distance between the neighbouring atoms. Thus, we are interested in the effects that take place at a certain distance from this vortex $r\gg{}R$.  

We assume that motion of electrons is described by the planar $2D$ Dirac equation 
\begin{equation}
H_D\Psi_\tau=i\partial_t\Psi_\tau,
\end{equation}
where the Dirac Hamiltonian operator 
\begin{align}
H_D&= -I\sigma_1 \left[i \partial_x - |e|A_x ( \vecr ) \right] - \tau_3\sigma_2 \left[ i \partial_y - |e|A_ y ( \vecr ) \right]
\label{Ha1}
\end{align}
includes the  gauge potentials $A_{x}(\vecr)$ and $A_{y}(\vecr)$ 
\footnote{Vector potential $\vec A$, instead of having an electromagnetic nature, may be due to certain defects in the graphene structure. In this case, we consider the so called pseudopotential description of our model of a  circular defect (for the description of a linear defect by pseudopotential note, in particular, that changes in the distance between the atoms and in the overlap between the different orbitals by strain or bending lead to changes in nearest--neighbor (NN) hopping or next--nearest--neighbor (NNN) hopping amplitude and this results in appearance of vector potentials $A_{x}(\vecr), A_{y}(\vecr)$ (this coupling may take the form of a gauge field with the matrix structure of the Pauli matrices, $\sigma_1$ and $\sigma_2$) in the Dirac Hamiltonian \cite{Geim0,Neto2} (see also Ref.~\onlinecite{line}).} 
with the charge $e=-|e|$, $\sigma_i$ as $2\times2$ Pauli-matrices in the pseudospin space
\footnote{The physical spin of electrons that is due to spatial rotation properties of the electron wave function has been neglected in our analysis, and the spinor nature of the wave function has its origin in the sublattice degrees of freedom called pseudospin.},
$\tau_3$ as the Pauli matrix in the valley subspace  with the eigenvalues $\tau=\pm 1$ for the two Fermi points $K,\,\,K'$, corresponding to valleys at the corners in the first Brillouin zone, $I$ is the unit matrix in the same space~\footnote{Our choice of signs in front of momentum and vector potential components of the Hamiltonian essentially corresponds to the conventions of \cite{Geim10,Neto2} (see also Ref.~\onlinecite{line}). It may differ from that of other papers due to different initial definitions adopted. However, the final results do not depend on it.}. 
Fermi velocity in what follows is supposed to be equal to unity with the corresponding choice of the units. Here  the 4--spinors in the 2D plane $\Psi_\tau (\vecr)$ ($\vecr=(x,\,\,y)$) consist of two 2--spinor components $\Psi_{i,\tau}$ ($i=~1,2$)
\begin{equation}
\Psi_\tau(\vecr)=
\begin{pmatrix}
\Psi_{1,\tau} \\
\Psi_{2,\tau} \\
\end{pmatrix}
\label{zhutt}
\end{equation}
describing electrons at the two $A,\,\,B$ sublattices ($i=~1,2$) with eigenvalues $\tau=\pm 1$ of $\tau_3$  for  two Fermi points $K,\,\,K'$. The expression Eq.~\eqref{Ha1} implies that the low-momentum expansion around the other Fermi point with $\tau \to -\tau$ gives rise to a time-reversed Hamiltonian. Note that the total effect of both valleys, as described in four--spinor notations (see Ref.~\onlinecite{Gus}, and references therein), 
respects time--reversal invariance. We shall use the appropriate 4--spinor notations \eqref{zhutt} in what follows to describe solutions of the problem more conveniently. 

Since our problem has cylindrical symmetry, we use polar coordinates $r,\,\varphi $ in the $xy$ plane, then the Dirac equation splits into a system of two equations  
\begin{equation}
\left\{
\begin{array}{l}
E\Psi_{1,\tau}+e^{-i\tau\varphi}(i\partial_{r}+\frac{\tau}{r}\partial_{\varphi}+i\tau{}|e|A_{\varphi})\Psi_{2,\tau}=0,\\
E\Psi_{2,\tau}+e^{i\tau\varphi}(i\partial_{r}-\frac{\tau}{r}\partial_{\varphi}-i\tau{}|e|A_{\varphi})\Psi_{1,\tau}=0.\\
\end{array}
\right.
\label{ha}
\end{equation}

One can see that there are relations between two components of the wave function with different valley indices $\tau=\pm1$, i.e. $\Psi_{1,\tau}=\Psi_{2,-\tau};\,\,\,\,\,\,\Psi_{2,\tau}=\Psi_{1,-\tau}$. According to these relations, the Dirac equation may be solved just for $\tau=+1$. For this reason, we shall omit ``+''sign in the subscripts in the following formulas, so $\Psi_1=\Psi_{1,+}$, etc. The role of other brunch of solutions with $\tau = -1$ will be considered later with the use of Appendix~\ref{Appendix}. We will search for solutions  in the following form: 
\begin{equation}
\left\{
\begin{array}{l}
\Psi_1(r,\varphi)=\e^{i(l-1)\varphi}\Psi_1(r), \\
\Psi_2(r,\varphi)=\e^{il\varphi}\Psi_2(r), \\
\end{array}
\right.
\end{equation}
where $l= \dots, -2,-1,0,+1,+2,\dots$ is the orbital quantum number. Let us introduce a dimensionless variable $\rho=r^2|e|B/2$, then the Dirac equations look like
\begin{equation}
\left\{
\begin{array}{l}
\Psi_1=-i\sqrt{\frac{\rho}\lambda}(\partial_{\rho}+\frac{l+\mu}{2\rho}+\frac12)\Psi_2, \\
\Psi_2=-i\sqrt{\frac{\rho}\lambda}(\partial_{\rho}-\frac{l+\mu-1}{2\rho}-\frac12)\Psi_1, \\
\end{array}
\right.
\label{Eq1}
\end{equation}
where instead of the energy $E$ we introduced a dimensionless parameter $\lambda={E^2\over 2|e|B}$ (we assume here that $B>0$; for situation with $\vec B$ antiparallel to $Oz$, see Appendix~\ref{Appendix}). One can substitute $\Psi_2$ from the second equation to the first one and vice versa to obtain the following equations
\begin{align}
&\left[\partial^2_{\rho}+\frac{1}{\rho}\partial_{\rho}-\frac14-\frac{(l+\mu-1)^2}{4\rho^2}+\frac{2\lambda-l-\mu}{2\rho}\right]\Psi_1(\rho)=0,
\label{EQ1}\\
&\left[\partial^2_{\rho}+\frac{1}{\rho}\partial_{\rho}-\frac14-\frac{(l+\mu)^2}{4\rho^2}+\frac{2\lambda-l-\mu+1}{2\rho}\right]\Psi_2(\rho)=0.
\label{EQ2}
\end{align}
The general solutions for these equations can be found (see also
Ref.~\onlinecite{Slob2})
\begin{align}
&\Psi_{1}(\rho) = C_1
\e^{-\rho/2}\rho^{|l+\mu-1|/2}\times\label{RE1}\\
&\times\Phi\left(\frac12(l+\mu+1+|l+\mu-1|-2\lambda),
1+|l+\mu-1|; \rho\right)
\notag\\
&+D_1
\e^{-\rho/2}\rho^{-|l+\mu-1|/2}\times\notag\\
&\times\Phi\left(\frac12(l+\mu+1-|l+\mu-1|-2\lambda),
1-|l+\mu-1|; \rho\right),
\notag\\
&\Psi_{2}(\rho) = C_2
\e^{-\rho/2}\rho^{|l+\mu|/2}\times\label{re2}\\
&\times\Phi\left(\frac12(l+\mu+|l+\mu|-2\lambda ),
1+|l+\mu|; \rho\right)\notag\\
&+D_2
\e^{-\rho/2}\rho^{-|l+\mu|/2}\times\notag\\
&\times\Phi\left(\frac12(l+\mu-|l+\mu|-2\lambda ),
1-|l+\mu|; \rho\right),
\nonumber
\end{align}
where $C_1$, $C_2$, $D_1$, $D_2$ are constants, related by the Dirac equations \eqref{ha}, and $\Phi(a,b;\rho)$ is the confluent hypergeometric function \cite{ryzhik}.

One can see that Eqs.~\eqref{EQ1}-\eqref{EQ2} and their solutions \eqref{RE1}-\eqref{re2} have the same form as for the pure homogeneous magnetic field \cite{SOKOLOV} (i.e. without the Aharonov--Bohm vortex), but for an additional summand  $\mu$, changing $l$ to $l+\mu$. Since the flux $\mu$ is not apriory an integer number, we introduce its fractional part $\tilde\mu$ according to
\begin{align}
&\mu=\mu_0+\tilde\mu, \\
&\mu_0=0,\pm1,\pm2\dots,\,\,0\le\tilde\mu<1. \nonumber
\end{align}

If $\rho\to\infty$ the confluent hypergeometric function behaves as $\e^{\rho}$, thus the hypergeometric series should be reduced to a Laguerre polynomial $L^{k}_{q}(\rho)$. To this end one should put the first arguments of the confluent hypergeometric functions in Eqs.~\eqref{RE1}--\eqref{re2} to be equal to the non-positive integer numbers. Unfortunately, it is impossible to reduce  both confluent hypergeometric functions that enter the solutions~\eqref{RE1}-\eqref{re2} of the Dirac Eqs.~\eqref{EQ1}-\eqref{EQ2} with coefficients $C$ and $D$ to a polynomial at the same time, since parameter $\lambda$ is proportional to the energy squared and therefore has the same value for both equations, but $|l+\mu-1|$, generally speaking, is not integer due to the presence of the non-integer term $\mu$. Thus, one has to put either $C$, or $D$ coefficient to be equal to zero. Here we take $D_1=D_2=0$ in order to keep the solution with a correct behaviour in the $\rho\to0$ limit \cite{Slob2}. Now, to reduce the hypergeometric series to a Laguerre polynomial we put the first argument of the confluent hypergeometric function in Eq.\eqref{re2} to be equal to the non-positive integer number $-s$ ($s=0,1,2,\,\dots$), which can be written in the form of the energy spectrum 
\begin{equation}
2\lambda=2s+l+\mu+|l+\mu|.
\label{lambda_}
\end{equation}
Then we go over to ``modified'' orbital quantum numbers
\begin{align}
\tilde l=l+\mu_0
\end{align}  
and obtain
\begin{equation}
\label{eq}
2\lambda=2s+l+\mu+|l+\mu|=2s+\tilde l +\tilde \mu+|\tilde l +\tilde \mu|,
\end{equation}
Consider two cases:

{\bf Let a)} $\tilde l\leq0$: Then, the energy spectrum \eqref{eq} does not depend on the flux $\mu$ and is given by $\lambda=n$, where $n=~0,1,2,\dots$ is the principal (Landau) quantum number. It is the well known result of appearance of quantized Landau levels in graphene in a magnetic field (see Ref.~\onlinecite{Katss} chapter 2.2), when the energy  $E=\sqrt{2|e|B\lambda}$ depends on $n$ like $E\sim\sqrt{n}$ \footnote {This is drastically different from the case of free non--relativistic electrons in a magnetic field, when $E=\hbar\omega(n+\frac12+\sigma),\,\,\sigma=\pm {1\over 2}$.}. The wave function in this case is given by the following relation  
\begin{align}
\Psi(\rho) &=
\begin{pmatrix}
\Psi_{1}(\rho) \\
\Psi_{2}(\rho) \\
\end{pmatrix}
\\
&=
\begin{pmatrix}
C_1\e^{i(l-1)\varphi}\e^{-\rho/2}
\rho^{-(\tilde l+\tilde\mu-1)/2}L_{n}^{-(\tilde l+\tilde\mu-1)}(\rho) \\
C_2\e^{il\varphi}\e^{-\rho/2}\rho^{-(\tilde
  l+\tilde\mu)/2}L_{n}^{-(\tilde l+\tilde\mu)}(\rho) \\
\end{pmatrix}.\nonumber
\label{state}
\end{align}
For this trivial case, we obtained the zero--energy modes with $n=0$, which are protected by the topology (see Ref.~\onlinecite{Katss} chapter 2.3). This means, that the electron trajectory does not capture the Aharonov--Bohm vortex and is affected only by the constant magnetic field. 

{\bf b) Let} $\tilde l>0$: Then, from Eq.~\eqref{eq} one has
\begin{equation}
\label{eq1}
2\lambda=2s+\tilde l +\tilde \mu+|\tilde l +\tilde \mu|=2(s+\tilde l +\tilde \mu),
\end{equation}
where $s=0,\,1,\,2,\dots$ is the radial quantum number. In this way, the confluent hypergeometric function transforms to the Laguerre polynomial $L_{s}^n$ (see, e.g. Ref.~\onlinecite{SOKOLOV}). Now the energy spectrum is given by the relation $\lambda=n+\tilde\mu$ in dimensionless terms, or \footnote{Similar result for (3+1)-D massive fermions  see Ref.\onlinecite{BGT1}.}
\begin{equation}
\label{Energy}
E^2=2|e|B(n+\tilde\mu),
\end{equation}
where the principle number is $n=s+\tilde l=s+l+\mu_0=1, 2, 3,\dots,$ where we recall that $\mu_0=0,\pm1,\pm2\dots.$ For this nontrivial case the wave function is given by 
\begin{align}
\Psi(\rho) &=
\begin{pmatrix}
\Psi_{1}(\rho) \\
\Psi_{2}(\rho) \\
\end{pmatrix}
\label{state}\\
&=
\begin{pmatrix}
K_1\e^{i(l-1)\varphi}\e^{-\rho/2}
\rho^{(\tilde l+\tilde\mu-1)/2}L_{s}^{\tilde l+\tilde\mu-1}(\rho) \\
K_2\e^{il\varphi}\e^{-\rho/2}\rho^{(\tilde
  l+\tilde\mu)/2}L_{s}^{\tilde l+\tilde\mu}(\rho) \\
\end{pmatrix}.\nonumber
\end{align}
It is important that in this case there are no zero energy states, which means that the topology is changed, in other words now the electron trajectory captures the Aharonov--Bohm vortex inside, which affects the energy spectrum. The energy spectrum is similar to the case of charged impurity on the two dimensional graphene surface (Ref.~\onlinecite{Katss} chapter 2.10), thus the Aharonov--Bohm vortex plays the role of the impurity or the defect.  

\section{Effective potential and vacuum current}
Here we focus our attention only on the nontrivial case, when electrons are affected both by a constant homogeneous magnetic field and the Aharonov--Bohm field. The effective potential of the model can be calculated using the Fock--Schwinger proper time method~\cite{fock1937collection} (see also Refs.~\onlinecite{Biet, Zhukovsky2012597, STZhB, bogoliubov1959introduction}) 
\begin{align}
V_{\rm eff}&=
\frac12\frac{1}{ST}\Tr\int_0^\infty\frac{dz}{z}\e^{-z[p_0^2+E^2]}\nonumber\\
&=\frac12\frac{1}{S}\sum
\limits_{q}
\frac{1}{2\pi}\int_0^\infty\frac{dz}{z}
\int_{-\infty}^{+\infty}{}dp_0\e^{-z(p^2_0+E^2)}\nonumber\\
&=\frac{1}{4S\pi}\sum\limits_{q}\int_0^\infty\frac{dz}{z}\sqrt{\frac{\pi}{z}}\e^{-zE^2},
\end{align}
where $ST$ is the (2+1)D volume, $S=\pi{}R^2_{0}$ is the area of the 2D graphene sample ($R_{0}$ is large but finite), and $E^2$ is the square of the energy \eqref{Energy}, i.e. the eigenvalue of $H_{D}^2$, for quantum states with numbers $q=n,s, \tau=\pm 1$, and $\epsilon=\pm 1$ as the sign of the energy $E=\epsilon|E|$. We shall further consider separately two possible cases: $\tau >0$ and $\tau< 0$:

{\bf Let a)} $\tau =+1$: Recall that according to $s+\tilde l=n$, where $s=0, 1, 2, ...$; $\tilde l=~1, 2, 3, ...$; we have $n=1, 2,3,\dots$ for Landau quantum number.  Thus summation over $\tilde l$ and $s$ can be replaced by summation over $n$ and $s$. According to \cite{SOKOLOV} one can  obtain the radius of the semiclassical circular electron trajectory $r_n=\sqrt {(2n+1)/|e|B}$, and consequently, the typical size of the localized states in Landau level, i.e. the so called magnetic length is equal to $r_B=\sqrt{1/|e|B}$. At the same time, the deviation of the center of the trajectory from the origin  in terms of quantum number $s$ is defined as $a=\sqrt{(2s+1)/|e|B}$. 

The  magnetic field strength in continuous regime can be taken to be of the order of $B \ge 1\,{\rm T}$ and in superconductor devices $\ge 10\,{\rm T}$, while in pulsed regime it can reach $\sim100\,{\rm T}$, the extremely strong field that could be obtained in the laboratory is $\sim1000\,{\rm T}$ in pulsed regime. 

Consider comparatively weak fields $B\sim 1\,{\rm T}$, when the magnetic length $r_{B}\le 30\, {\rm nm}$. The radius of the Aharonov--Bohm vortex in graphene is comparable to the distance between the neighbouring atoms and can be approximated by $R=1\,{\rm nm}$. Then we may assume that $r_{B} \gg R$. For instance, if $R=1\,{\rm nm}$ and $r_{B}=30\,{\rm nm}$ the ratio $(R/r_B)^2\approx 10^{-3}$. At the same time the size of the graphene sheet may be $R_0\ge 10 R$. 

The energy spectrum depends only on $n$ and does not include quantum number $s$, which means, in classical terms, that the center of the circular trajectory of a particle can not be fixed, and hence summation over this number should be made with regard to this degeneracy. For the classical electron trajectory to "embrace" the circular vortex of radius $R$, we have a natural geometrical condition $r_n\geq R+a$, which is equally true for $R<a$, as well as for $R=a$, or $R>a$. In this case $\sqrt{2n+1}\ge\sqrt{|e|B}R+\sqrt{2s+1}$ and in weak fields $\sqrt{|e|B}R\ll 1$ we have $n\ge s$. For the realistic situation with magnetic fields as described above, we arrive at the condition for summation over $n$ and $s$ (recall that under the condition for $\tilde l >0$ the trajectory of the particle "embrace"s the vortex, and the value $n=0$ is excluded from the spectrum) 
\begin{equation}
n \ge s+1,\,\, 0 \le s <s_{\rm max},
\end{equation}
where $s_{\rm max}\approx |e|B R_0^2/2$. Then summation over  $n$ is made, assuming that the upper limit under the condition $|e|B\ll~1/R^2$ can
be extended to infinity
\begin{equation}
\sum\limits_{n=s+1}^{+\infty}\e^{-2|e|Bzn}=\frac{\e^{-2|e|Bz(s+1)}}{1-\e^{-2|e|Bz}},
\label{n1}
\end{equation}
and hence summation over $s$ results in
\begin{equation}
\sum\limits_{s=0}^{s_{\rm max}}\sum\limits_{n=s+1}^{\infty}\e^{-2|e|Bzn}
= \frac{\e^{-2|e|Bz}}{1-\e^{-2|e|Bz}}
\frac{1-\e^{-2|e|Bz(s_{\rm max}+1)}}{1-\e^{-2|e|Bz}}.
\label{s}
\end{equation}
In the problem without a vortex for a uniform plane with a homogeneous magnetic field perpendicular to the plane this limitation $n>s$ for summation over $n$ should be omitted and the contribution of $n=0$ with consideration for only one spin state for $n=0$ should be included. Then, instead of \eqref{n1} one can get 
\begin{align}
\label{adding}
2\left(\sum\limits_{n=1}^{+\infty}\e^{-2|e|Bzn}+\frac{1}{2}\right)=\frac{2\e^{-2|e|Bz}}{1-\e^{-2|e|Bz}}+1=\coth |e|Bz.
\end{align}
Here factor 2 accounts for two spin orientations in the state with $n\ne 0$, while in the ground state $n=0$ only one spin orientation is possible and this gives  an additional second term in \eqref{adding}.
Summation over $s$ gives the degeneracy factor
\begin{align}
\label{degeneracy}
\sum\limits_{s=0}^{s_{{\rm max}}} 1\,\,=s_{{\rm max}}+1\approx |e|BR_0^2/2.
\end{align}
Thus we may arrive, after appropriate subtraction and renormalization, at just what we have for 4D space-time effective Lagrangian, i.e. the well known Heisenberg--Euler formula (see, e.g., \cite{Schwinger_sr,STZhB}) -- in a magnetic field 
\begin{align}
{\cal L_{\rm eff}}=-{1\over 8\pi^2}\int\limits_0^\infty {dz\over
 z}
\e^{-zm^2}\left [\frac{|e|B}{z}\coth (|e|zB)- {1\over z^2}- {\e^2B^2\over
      3}\right ].
\end{align}

In this paper we consider the situation with an Aharonov--Bohm vortex, described by formula \eqref{s}, and we use the fundamental restriction $\tilde l >0$ for the wave-function, which means that the semiclassical trajectory of the particle should "embrace" the vortex. Moreover, the value $n=0$ is excluded from the spectrum to avoid particle penetration into the vortex. 

One should note, that with these conditions taken into account, the results for the effective potential and for the induced current (which follows) can not go over to the corresponding limiting formulas with $B\to 0$.

Under these restrictions, we obtain the effective potential that takes into account both the Aharonov--Bohm vortex and the homogeneous magnetic field.
\begin{align} 
V_{\rm eff} &=
\frac{1}{4S\pi}\sum\limits_{n,s}\int_{\frac {1}{\Lambda ^2}}^\infty\frac{dz}{z}\sqrt{\frac{\pi}{z}}\e^{-zE^2}\label{int}\\
&=\frac{1}{4S\pi}
\int_{\frac {1}{\Lambda ^2}}^\infty\frac{dz}{z}\sqrt{\frac{\pi}{z}}
\frac{(1-\e^{-2eBzs_{\rm max}})\e^{-2eBz(\tilde\mu+1)}}{(1-\e^{-2eBz})^2}
.\nonumber
\end{align}
This integral diverges at the lower limit $z\to 0$, thus we introduced the cutoff parameter $\Lambda^{-2}$, such that $|e|B\Lambda^{-2}\ll~1$. The main contribution to this integral is given by $ z|e|B\ll 1$, then $z|e|Bs_{\rm max}\sim |e|Bs_{\rm max}\Lambda^{-2}\ll 1$. We study the real material that has a periodical atomic structure and according to the Debye theory of  solids we should introduce the physical restriction on the wavelength and in consequence on the value of the proper time lower limit $\Lambda^{-2}$ that will regularize the result. This restriction should be based on the structure of the material, i.e. it should be related to the characteristic length of the model. Parameter $\Lambda^{-2}$ has the dimension of $[{\rm   Length}^2]$ and one may put it equal to the area of the Aharonov--Bohm vortex in graphene $\Lambda^{-2}=\pi{}R^2$. Then $z|e|Bs_{\rm max}\sim |e|Bs_{\rm max}\Lambda^{-2}\sim (|e|B)^2 R_0^2\pi R^2/2$ and if $(R/r_B)^2\approx 10^{-3}$, for $R_0\ge 10 R$ we have $z|e|Bs_{\rm max}\sim 10^{-6}(R_0/R)^2\sim 10^{-4}\ll 1$ and  
\begin{align}
\frac{1-\e^{-2|e|Bzs_{\rm max}}}{(1-\e^{-2|e|Bz})^2}\approx 
\frac{s_{\rm max}}{2|e|Bz}=\frac{R_0^2}{4z}.
\end{align}
Thus, the degeneracy factor, which is proportional to $R_0^2$, is cancelled by $S$ in the denominator of Eq.\eqref{int}, and the result is  
\begin{align}
\label{veff}
V_{\rm eff} &=\frac{1}{4S\pi}\int_{\Lambda^{-2}}^\infty 
\frac{dz}{z}\sqrt{\frac{\pi}{z}}\,\frac{s_{\rm max}}{2|e|Bz}\,
\e^{-2|e|Bz(\tilde\mu+1)}
\nonumber\\
&= \frac{1}{16\pi^2}\int_{\Lambda^{-2}}^\infty
\frac{dz}{z^2}\sqrt{\frac{\pi}{z}}\,\e^{-2|e|Bz(\tilde\mu+1)}.
\end{align}
In fact, we may  apply the restriction $n\ge 1$ and approximately use Eq.~\eqref{n1} for summation starting with $n=1$ 
\begin{equation}
\sum\limits_{n=1}^{+\infty}\e^{-2|e|Bzn}=\frac{\e^{-2|e|Bz}}{1-\e^{-2|e|Bz}},
\label{n}
\end{equation}
and then sum over $s$ using summation formula \eqref{degeneracy} and we arrive  at the same Eq.~\eqref{veff}, thus justifying the validity of our above made approximations.
 
The integral in Eq.~\eqref{veff} can be computed exactly
\begin{align}
\label{veff1}
V_{\rm eff}=\,\,
&\frac{\sqrt{\pi}}{16\pi^2}\left(\frac43\sqrt{\pi}\big(2|e|B(\tilde\mu+1)\big)^{3/2}{\rm erf}\big(\sqrt{2|e|Bz(\tilde\mu+1)}\big) \right.\nonumber\\
&+\left.\left.\frac{8|e|Bz(\tilde\mu+1)-2}{3z^{3/2}}\e^{-2|e|Bz(\tilde\mu+1)}\right)\right|^{+\infty}_{\Lambda^{-2}},
\end{align}
where ${\rm erf}$ is the Error Function. Considering ${\rm erf}(0)=~0$, ${\rm erf}(+\infty)=~1$ and $|e|B\Lambda^{-2}\ll~1$, one obtains 
\begin{align}
V_{\rm eff}=\,\,&
\frac{\big(2|e|B(\tilde\mu+1)\big)^{3/2}}{12\pi}-\notag\\
&\frac{\big(4|e|B(\tilde\mu+1)-\Lambda ^2\big)\Lambda}{24\pi^{3/2}}\e^{-2|e|B(\tilde\mu+1)/\Lambda^2}.
\end{align}
Subtracting the term that does not depend on the magnetic field $B$, in the limit  $|e|B\ll \Lambda^2$ putting $\Lambda^{-2}=\pi{}R^2$, we have
\begin{equation}
\label{eq1}
V_{\rm eff}= - {|e|B(\tilde \mu+1)\over 4\pi^{2}R}+\frac{\big(2|e|B(\tilde\mu+1)\big)^{3/2}}{12\pi}.
\end{equation}
As an illustration of the vacuum structure of the model we calculate the induced current
\begin{align}
J_\varphi={\partial V_{\rm eff} \over\partial A_\varphi^{AB}},
\end{align}
where $A_\varphi^{AB}$ is the Aharonov--Bohm potential \eqref{AB}, defined for $r\ge R$, so $A_\varphi^{AB}=\frac{\mu}{r|e|}$. Recall that we considered in this part of the article the contribution of the branch of the spectrum with $\tau>0$. In order to further distinguish this contribution we now call the corresponding part of the current $J_\varphi^{(+)}$. We have from \eqref{veff}
\begin{align}
\label{result}
&J_\varphi^{(+)}=
|e|r\frac{\partial{}V_{\rm eff}}{\partial\tilde\mu}=
-\frac{e^2Br}{8\pi^2}\int_{\frac {1}{\Lambda^2}}^\infty\frac{dz}{z}
\sqrt{\frac{\pi}{z}}\e^{-2|e|Bz(\tilde\mu+1)}=\notag\\
&-\frac{\e^2Br\sqrt{\pi}}{4\pi^2}\left(\sqrt{\pi}\sqrt{2|e|B(\tilde\mu+1)}\,\,{\rm erf}\big(\sqrt{z2|e|B(\tilde\mu+1)}\big)\right.\notag\\
&\hspace{1.62cm}+\left.\left.\frac{\e^{-2|e|Bz(\tilde\mu+1)}}{\sqrt{z}}\right)\right|^{\infty}_{\Lambda^{-2}},
\end{align}
or considering ${\rm erf}(0)=~0$, ${\rm erf}(+\infty)=1$, with the assumption that $|e|B\Lambda^{-2}\ll~1$, we have
\begin{align}
\label{curr}
&J_\varphi^{(+)}=-\frac{e^2B}{4\pi^2}\frac{r}{R}\left(1-\pi R\sqrt{2|e|B(\tilde\mu+1)}\right),\nonumber\\
&{\rm for}\,\,R_0\ge r\ge R,\,\,R\sqrt{|e|B}\ll 1.
\end{align}

{\bf Let b)} $\tau=-1$: Now we consider contribution of the $\tau=-1$ brunch of the spectrum. According to Eq.~\eqref{Energy_} from Appendix~\ref{Appendix} we have for the energy now 
\begin{equation}
\label{Energy_1}
E^2=2|e|B(n-\tilde\mu), 
\end{equation}
and hence
\begin{equation}
\label{eq1}
V_{\rm eff}^{(-)}= - {|e|B(1-\tilde \mu)\over 4\pi^{2}R}+\frac{\big(2|e|B(1-\tilde\mu)\big)^{3/2}}{12\pi}.
\end{equation}
In this case the solution for the induced current will look like 
\begin{align}
\label{curr_}
&J_\varphi^{(-)}=\frac{e^2B}{4\pi^2}\frac{r}{R}\left(1-\pi R\sqrt{2|e|B(1-\tilde\mu)}\right),\nonumber\\
&{\rm for}\,\,R_0\ge r\ge R,\,\,R\sqrt{|e|B}\ll 1.
\end{align} 
The total current is the sum $J_\varphi^{tot}=J_\varphi^{(+)}+J_\varphi^{(-)}$. Summing Eqs.~\eqref{curr} and \eqref{curr_}, with the divergent at $R\to 0$ contributions cancelling each other, one can see that 
\begin{align}
J_\varphi^{tot}= \frac{e^2Br}{4\pi}(\sqrt{2|e|B(\tilde\mu+1)}-\sqrt{2|e|B(1-\tilde\mu)}).
\end{align}

An interesting question arises: what happens when the Aharonov--Bohm flux vanishes. Since in our calculations, we assumed that only those solutions provide nonzero contribution to the current that ``embrace'' the solenoid, there is no simple way to go to the limit of vanishing $\tilde\mu\to~0$ starting just with Eq.~\eqref{curr}. Indeed, for this purpose we have to take the sum
\begin{align}
J_\varphi^{tot}(\tilde\mu=0)=J_\varphi^{(+)}(\tilde\mu\to
  0)+J_\varphi^{(-)}(\tilde\mu\to 0),
\end{align}
which gives $J^{tot}_\varphi(\tilde\mu=0)=0$ (with the divergent at $R\to 0$ contributions (see Eq.~\eqref{curr}) cancelling each other). By the way, we see that the following symmetry takes place
\begin{align} 
J_\varphi^{tot}(\tilde\mu)=-J_\varphi^{tot}(-\tilde\mu),
\end{align}
which is in agreement with Jackiw--Milstein conclusion \cite{JaMi, PhysRevB.83.075420} for the problem without homogeneous magnetic field. At the same time, the presence of the preferred direction in this problem (the magnetic field) diminishes parity property with $z\to -z$ and as a consequence there is no symmetry $l\to -l$, and contrary to Jackiw--Milstein conclusion valid for only Aharonov--Bohm flux participating, there is no symmetry $|\tilde\mu|\to|\tilde\mu|-1$ in this problem.  

One remark is in order. In this problem, there is a preferred direction along the magnetic field, which we chose to be parallel to $Oz$ axis, i.e. $\vec B =(0,\,0\,B)$ with $B>0$. At the same time we agreed to take the Aharonov--Bohm flux $\mu$ to be positive, when it is also directed along the $Oz$ axis and thus is parallel to $\vec B$. In the situation, when the magnetic field points in the direction opposite to $Oz$, $\vec B=(0,\,0\,-B)$, while $\mu$ is still directed along $Oz$, solutions of the Dirac equation change and for the energy we have (see Appendix~\ref{Appendix})
\begin{equation}
\label{Energy_12}
E^2=2|e|B(n-\tilde\mu). 
\end{equation}
It is evident that in the inverse situation with the magnetic field still pointing in the $z-$direction, but the $\mu-$flux taken to be positive when it is antiparallel to $Oz$ and, hence, to $\vec B$, the solution \eqref{curr_} is also valid.

\section{Electric current in strong magnetic field}
It is also possible to determine not only the vacuum current, but also the current of the real electrons in our model in the presence of a magnetic field. Description of the wave function of the real electrons as a superposition of the bare states \eqref{state} is a rather difficult procedure, because it is impossible to determine exactly in what state every electron is located. Nevertheless, there is one particular case, where it is possible to determine such wave package. Electrons under the strong magnetic field lay on the lowest energy level $n=0$, which is infinitely degenerate in the quantum number $l\leqslant0$. The energy spectrum in the limit of strong magnetic field corresponds to the trivial case in the absence of the vortex and is given by $\lambda = n$. Then, electronic state can be described by the wave function 
\begin{align}
\Psi^{\rm HFL}(\rho)
=\sum\limits_{l\leqslant0}C_{l}
\begin{pmatrix}
\e^{i(l-1)\varphi}\e^{-\rho/2}
\rho^{-(\tilde l+\tilde\mu-1)/2} \\
\e^{il\varphi}\e^{-\rho/2}\rho^{-(\tilde
  l+\tilde\mu)/2} \\
\end{pmatrix},
\end{align}
because $L_0^{a}(\rho)=1$. Then, the electric current is given by the following equation
\begin{align}
j_{x} =\,\,
&\Psi^{\dagger}\sigma_{x}\Psi \nonumber\\
=\,\, &e^{-\rho}\rho^{\frac12-\tilde\mu}\left[\sum\limits_{l\leqslant0}C_{l}
e^{-il\varphi}\rho^{-\tilde{l}/2}\right]
\left[\sum\limits_{l\leqslant0}C_{l}
e^{il\varphi}\rho^{-\tilde{l}/2}\right]e^{i\varphi}+\nonumber\\
&e^{-\rho}\rho^{\frac12-\tilde\mu}\left[\sum\limits_{l\leqslant0}C_{l}
e^{-il\varphi}\rho^{-\tilde{l}/2}\right]
\left[\sum\limits_{l\leqslant0}C_{l}
e^{il\varphi}\rho^{-\tilde{l}/2}\right]e^{-i\varphi}\nonumber\\
=\,\,&J^0_{\rm r}\cos\varphi,\nonumber\\
j_{y}
=\,\,&J^0_{\rm r}\sin\varphi,
\end{align}
which means, that electric current of magnitude $J^0_{\rm r}$ is directed along the Aharonov--Bohm vortex that plays the role of an impurity in this problem.

\section{Conclusions}
In this paper, the electronic wave functions and energy spectrum for the graphene model with the superposition of a constant magnetic field and an Aharonov--Bohm vortex were obtained. For the non-positive values of the ``modified'' orbital quantum number $\tilde l\leq0$ the energy spectrum does not depend on the magnetic flux of the Aharonov--Bohm field, therefore the trajectory of electrons does not capture the magnetic vortex. It is important that for this case the energy spectrum is degenerate, which allows us to calculate electronic current for the strong magnetic field limit. This current is directed strictly along the vortex, which was demonstrated in the last Section. For the positive values of $\tilde l>0$, the energy spectrum explicitly depends on the Aharonov--Bohm flux and the magnetic vortex plays the role of a charged defect. Furthermore, using the effective potential method, the expressions for the vacuum current in the weak and strong field limits were obtained.   

\appendix
\section{Energy spectrum for the second valley and the opposite orientation of the magnetic field}
\label{Appendix}
{\bf a) Energy spectrum for the second valley:}\\
Let us consider the case of $\tau =-1$. Then 
\begin{align}
\left\{
\begin{array}{l}
E\Psi_{1,-}+e^{+i\varphi}(i\partial_{r}-\frac{1}{r}\partial_{\varphi}-i{}|e|A_{\varphi})\Psi_{2,-}=0,\\
E\Psi_{2,-}+e^{-i\varphi}(i\partial_{r}+\frac{1}{r}\partial_{\varphi}+i{}|e|A_{\varphi})\Psi_{1,-}=0.\\
\end{array}
\right.
\label{ha_}
\end{align}
Now
\begin{align}
\left\{
\begin{array}{l}
\Psi_{1,-}(r,\varphi)=\e^{il\varphi}\Psi_{1,-}(r), \\
\Psi_{2,-}(r,\varphi)=\e^{i(l-1)\varphi}\Psi_{2,-}(r),
\end{array}
\right.
\end{align}
where $l= \dots, -2,-1,0,+1,+2,\dots$ is the orbital quantum
number. Recall $\Phi=2\pi\mu/|e|$. 
Then
\begin{align}
A_{\varphi}=\frac{\mu/|e|}{{}r}+\frac12 Br,
\end{align}
and
\begin{align}
\left\{
\begin{array}{l}
E\Psi_{1,-}(r)+(i\partial_{r}-\frac{1}{r}i(l-1)-i{}|e|A_{\varphi})\Psi_{2,-}(r)=0,\\
E\Psi_{2,-}+(i\partial_{r}+\frac{1}{r}il+i{}|e|A_{\varphi})\Psi_{1,-}=0,
\end{array}
\right.
\label{ha_}
\end{align}
or
\begin{align}
\left\{
\begin{array}{l}
E\Psi_{1,-}(r)+(i\partial_{r}-\frac{1}{r}i(l-1)-i{\mu\over r}-\frac i2 Br)\Psi_{2,-}(r)=0,\\
E\Psi_{2,-}(r)+(i\partial_{r}+\frac{1}{r}il+i{\mu\over r}+\frac i2 Br)\Psi_{1,-}(r)=0,
\end{array}
\right.
\label{ha_0}
\end{align}
and finally
\begin{align}
\left\{
\begin{array}{l}
\Psi_{1,-}=-i\sqrt{\frac{\rho}\lambda}(\partial_{\rho}-\frac{l-\mu-1}{2\rho}-\frac{1}{2})\Psi_{2,-}, \\
\Psi_{2,-}=-i\sqrt{\frac{\rho}\lambda}(\partial_{\rho}+\frac{l-\mu}{2\rho}+\frac{1}{2})\Psi_{1,-},
\end{array}
\right.
\label{Eq__00}
\end{align}
where $\gamma=|e|B/2$ and $\rho=\gamma r^2$. 
This is equivalent to replacement $\mu\to -\mu$ and  $
\Psi_{1,-}\to\Psi_{2,-}$ and  $
\Psi_{2,-}\to\Psi_{1,-}$ in Eq.~\eqref{Eq1}, leaving the energy spectrum \eqref{Energy} unchanged
but for replacement $\tilde\mu\to-\tilde\mu$
\begin{align}
\label{Energy_}
E^2=2|e|B(n-\tilde\mu). 
\end{align}

{\bf b) Energy spectrum for the case of magnetic field $\vec B$ antiparallel $Oz$:}\\ 
In Eqs.~\eqref{ha}, \eqref{Eq1} the following transformations can be made
\begin{align}
\partial_r-{l-1\over r}-\gamma r &= \sqrt{\gamma \rho}\left[
  2 \partial_\rho-1 -{l-1\over \rho}\right],\\
\partial_r+{l\over r}+\gamma r &= \sqrt{\gamma \rho}\left[
  2 \partial_\rho+1 +{l\over \rho}\right],
\end{align} 
where $\gamma=|e|B/2$ and $\rho=\gamma r^2$.
Let us suppose that $\vec B$ points in the opposite direction ($\vec B
=(0,\,0\,-B)$). This leads to change in the signs in front of ``1''
in the above equations  and the following replacements $\e^{i(l-1)\varphi}\to~\e^{-il\varphi}$
and $\e^{il\varphi}\to\e^{-i(l-1)\varphi}$.  Then  instead of
\eqref{Eq1} we obtain equations
\begin{align}
\left\{
\begin{array}{l}
\Psi_1=-i\sqrt{\frac{\rho}\lambda}(\partial_{\rho}-\frac{l-\mu-1}{2\rho}-\frac12)\Psi_2, \\
\Psi_2=-i\sqrt{\frac{\rho}\lambda}(\partial_{\rho}+\frac{l-\mu}{2\rho}+\frac12)\Psi_1, \\
\end{array}
\right.
\label{Eq_}
\end{align}
which is equivalent to replacement $\mu\to -\mu$ and  $
\Psi_1\to\Psi_2$ and  $
\Psi_2\to\Psi_1$, leaving the energy spectrum \eqref{Energy} unchanged
but for replacement $\tilde\mu\to-\tilde\mu$
\begin{align}
\label{Energy_}
E^2=2|e|B(n-\tilde\mu). 
\end{align}

\acknowledgments

The authors would like to thank C. Dutreix, S. Brener and
A.E. Lobanov for valuable discussions.

\bibliography{Circle}

\end {document}